\documentstyle[aps,aps10,a4,graphicx,amsfonts]{revtex}
\title{Transient QED effects in absorbing dielectrics}
\author{Martijn Wubs\thanks{E-mail: wubs@phys.uva.nl} \\
{\em \scriptsize Van der Waals-Zeeman Instituut, Universiteit van
Amsterdam, \\ Valckenierstraat 65, 1018 XE  Amsterdam, The
Netherlands}\and
\\ \mbox{L.G. Suttorp} \\ {\em \scriptsize Instituut voor
Theoretische Fysica, Universiteit van Amsterdam, \\
Valckenierstraat 65,  1018 XE Amsterdam, The Netherlands}}
\date{Submitted on 4 October 2000; \\
      to appear in Phys. Rev. A on 1 April 2001 }

\begin{document}
\maketitle

\begin{abstract}
The spontaneous emission rate of a radiating atom reaches its time-independent
equilibrium value after an initial transient regime. In this paper we consider
the associated relaxation effects of the spontaneous decay rate of atoms in
dispersive and absorbing dielectric media for atomic transition frequencies
near material resonances. A quantum mechanical description of such media is
furnished by a damped-polariton model, in which absorption is taken into
account through coupling to a bath. We show how all field and matter operators
in this theory can be expressed in terms of the bath operators at an initial
time. The consistency of these solutions for the field and matter operators
are found to depend on the validity of certain velocity sum rules. The
transient effects in the spontaneous decay rate are studied with the help of
several specific models for the dielectric constant, which are shown to follow
from the general theory by adopting particular forms of the bath coupling
constant.

\vspace{\baselineskip}
\noindent
PACS number(s): 42.50.Ct,71.36+c,03.70.+k
\end{abstract}

\section{Introduction}\label{Introduction}

The rate and the spectral and spatial characteristics of the spontaneous decay
of an atom depend on the properties of the atom and of the radiation field,
and on the interaction between them. The radiation field changes by the
presence of other matter \cite{Purcell46}. One can try and manipulate the
emission properties once the influence of this medium is understood.

In quantum optics of linear dielectrics, one tries to describe the material
medium in an effective way with the help of the classical dielectric function
$\varepsilon({\bf r}, \omega)$, which in general is a complex function of both
position and frequency and in this full generality describes the propagation
and loss of light at each point in the dielectric. Sometimes it is possible to
neglect  the spatial variations (including local field effects), dispersion
and losses altogether. The spontaneous emission rate of an atom in such a
simple dielectric is the refractive index $n$ of the medium times the rate
$\Gamma_{0}$ in vacuum \cite{diBartolo68} -- \cite{KhosraviL91}.

The situation becomes more complicated when material dispersion has to be
taken into account \cite{NienhuisA76} -- \cite{BarnettHLM96}. Since the
Kramers-Kronig relations tell that dispersion and loss always come together
(be it not always at the same frequencies), one should like to include losses
as well in order to describe all frequencies in one theory. The damped
polariton model \cite{KnollL92} -- \cite{HoK93} provides us with such a
microscopic theory. From that theory it was shown that the radiative
spontaneous emission rate equals $\Gamma_{0}$ times the real part of the
refractive index at the transition frequency \cite{BarnettHL92}.

The quantum mechanical treatment of dissipative systems is more complicated
than the classical one, because of the extra requirement that equal-time
commutation relations do not change over time \cite{Gardiner91,MandelW95}.
Based on the damped-polariton model and on the fluctuation-dissipation
theorem, phenomenological quantization theories were constructed that meet
these requirements. In these theories, the dielectric function is an input
function and the Maxwell field operators satisfy quantum Langevin equations
with both loss and quantum noise terms \cite{GrunerW95,MatloobL95}. With the
use of a Green-function approach, the phenomenological quantization theories
have been generalized to inhomogeneous dielectrics, first for multilayer
systems and later for general $\varepsilon({\bf r}, \omega)$ \cite{GrunerW96}
-- \cite{GrunerWgen}. Field commutation relations turn out only to depend on
the analytical properties of the Green function. However, the calculation of
spontaneous emission inside such a medium would involve the actual computation
of the Green function, which for general $\varepsilon({\bf r}, \omega)$ is not
easy.

A special case of the former theories is the quantum optical description of
inhomogeneous systems at frequencies where both dispersion and losses can be
neglected. Then a description in terms of modes is possible, where the mode
functions are harmonic solutions of the classical wave equation featuring a
position-dependent dielectric ``constant'' $\varepsilon({\bf r})$
\cite{Knollvv}. This encompasses the now theoretically and experimentally very
active research area of the so-called photonic crystals \cite{photfirst},
where a periodic modulation of the refractive index at the scale of the
wavelength of light can drastically modify the mode structure compared to
vacuum. By increasing the refractive-index contrast, even a photonic bandgap
can open up, giving rise to a frequency interval for which waves cannot travel
in the crystal in any direction, so that spontaneous emission would be
inhibited completely. Until now, such a bandgap has not been found
conclusively in the optical regime \cite{photmeas}. It has been proposed to
look for frequencies close to material resonances, where refractive indices
can be quite substantially higher or lower than 1 \cite{Moroz99}. 

Interesting new effects have been predicted for bandgap systems, such as
photon-atom bound states   and non-exponential spontaneous decay at the edges
of the gap \cite{JohnW91}. A current debate is whether the Weisskopf-Wigner
approximation can be used in the calculation of spontaneous emission near an
edge of a photonic bandgap. This question seems to depend strongly on the
analytic or singular behavior of the density of states at the edges of the
gap, which has recently been calculated for  face-centered cubic and
diamond-like  crystal structures \cite{LiZ00}. If near the edge of the bandgap
a large part of the modes has a cavity-like structure, producing nonzero
dwell-times near the emitting atom, then an emitted photon has a nonzero
probability of being reabsorbed, which would give Rabi-like oscillations of
the atomic population that are missed in the Weisskopf-Wigner approximation.

Non-exponential decay can also be caused by the interference of possible
decay-channels: for short times after the excitation of the atom, a larger
frequency interval of the medium states plays a part in the decay process than
for later times. Ultimately, only the refractive index at the atomic
transition frequency plays a role, all in concordance with the energy-time
uncertainty relation. This interference process already happens for
spontaneous emission in vacuum. However, when the medium has a strong jump in
the density of states around the atomic transition frequency, the interference
effect will change substantially.

To separate the latter cause of non-exponential decay from the former, it is
interesting to consider the spontaneous emission inside homogeneous lossy
dielectrics with strong and narrow material resonances, where the density of
states can also change very rapidly. Here all states correspond to simple
plane wave modes, so that real reabsorption processes do not play a role. In
this article, we use the damped-polariton model formulated by Huttner and
Barnett \cite{HuttnerB92a,HuttnerB92b} to study the interference effects of
spontaneous emission. If absorption is neglected in the damped-polariton
model, then we are left with the Hopfield model of a dielectric
\cite{HuttnerBB91,Hopfield58}, which has a frequency bandgap inside which the
refractive index is purely imaginary. The analogy between this polariton band
gap system and photonic crystals was drawn in \cite{RupasovS97}.

The organization of the paper is as follows: in section \ref{model} we
introduce the theory and solve its equations of motion using Laplace
transformations. In section \ref{sumrules} we show that the consistency of our
solutions depends on the validity of a number of velocity sum rules, which are
then proved. In section \ref{longtime}, we find that for long times all field
operators can be expressed in terms of the initial bath operators, and we give
an interpretation of the result. We also show how to relate the result to
phenomenological quantization theories. Before we can discuss transient
effects of spontaneous emission in section \ref{SE}, we discuss in section
\ref{modeleps} the Lorentz oscillator model and the point scattering model. We
show how both these models can be found from the damped-polariton theory by
choosing a suitable coupling to the bath. The paper ends with a discussion of
the results and with conclusions in section \ref{conc}.

\section{The model and solutions of the equations of motion}\label{model} The
damped-polariton theory describes the interaction of light with an absorbing
homogeneous medium. The coupling of the matter to a frequency continuum is the
cause of the light absorption. The continuum could be a phonon bath or
something else, but for the moment that is not specified: it is a collection
of harmonic oscillators with a frequency-dependent coupling to the matter
fields. Since the medium is homogeneous, the dynamics can be separated into a
transverse and a longitudinal part. In this article we concentrate on the
transverse excitations as described by the following Hamiltonian
\cite{HuttnerB92a,HuttnerB92b}:
\begin{equation}\label{HHB}
H = H_{\rm em} + H_{\rm mat} +H_{\rm bath} +H_{\rm int},
\end{equation}
with
\begin{eqnarray}\label{HHHH}
H_{\rm em}  &=& \int \mbox{d}^{3}k \; \hbar \tilde k c\;
a^{\dagger}(\lambda, {\bf k}, t)
 a(\lambda, {\bf k}, t), \\
H_{\rm mat}  &=& \int \mbox{d}^{3}k\; \hbar \tilde \omega_{0}\;
b^{\dagger}(\lambda, {\bf k}, t) b(\lambda, {\bf k}, t), \\
H_{\rm bath}  &=& \int \mbox{d}^{3}k \int_{0}^{\infty}d\omega \;\hbar
\omega \;
 b_{\omega}^{\dagger}(\lambda, {\bf k}, t)
 b_{\omega}(\lambda, {\bf k}, t), \\
H_{\rm int}  &=& \frac{1}{2}\int \mbox{d}^{3}k\;
\int_{0}^{\infty}d\omega\;\hbar V(\omega) \left[  b(\lambda, {\bf
k}, t) +
 b^{\dagger}(\lambda, -{\bf k}, t)\right]
\left[ b_{\omega}^{\dagger}(\lambda, {\bf k}, t) +
 b_{\omega}(\lambda, -{\bf k}, t)\right] \nonumber \\
& & + \frac{i}{2} \int
\mbox{d}^{3}k\;\hbar\omega_{c}\sqrt{\frac{\tilde\omega_{0}}{\tilde
k c}} \left[ a(\lambda, {\bf k}, t) +
 a^{\dagger}(\lambda, -{\bf k}, t)\right]
\left[ b^{\dagger}(\lambda, {\bf k}, t) -
 b(\lambda, -{\bf k}, t) \right].
\end{eqnarray}
We use the same notations as in \cite{HuttnerB92b}. In particular, $\tilde k
c$ stands for $\sqrt{k^{2}c^{2} + \omega_{c}^{2}}$, where the frequency
$\omega_{c}$ equals $\alpha/\sqrt{\rho\varepsilon_{0}}$, with $\alpha$ the
coupling constant between field and matter, and $\rho$ the density. The
resonance frequency $\omega_{0}$ of the polarization field is renormalized to
$\tilde\omega_{0}$, which is the positive-frequency solution of
\begin{equation}\label{tildeomnul}
\tilde\omega_{0}^{2} = \omega_{0}^{2} +
\tilde\omega_{0}\int_{0}^{\infty}\mbox{d}\omega\;V^{2}(\omega)/\omega.
\end{equation}
The $k$-integrals in the Hamiltonian are understood to also denote a summation
over the two transverse polarization directions labeled by $\lambda$. The
creation and annihilation operators satisfy standard bosonic commutation
relations. The Heisenberg equations of motion for the bath annihilation
operators are:
\begin{equation}\label{eqmotbath}
 \dot b_{\omega}(\lambda, {\bf k},t)  =  -\frac{i
}{2} V(\omega) \left[ b(\lambda, {\bf k},t) + b^{\dagger}(\lambda,
{\bf -k},t)\right] -i\omega b_{\omega}(\lambda, {\bf k},t),
\end{equation}
and similarly for the creation operators. In the following we drop the
$(\lambda, {\bf k})$-labels. We  solve implicitly for the bath variables, as
was done in \cite{DutraF98} in a classical treatment of the model:
\begin{equation}\label{bathsolv}
b_{\omega}(t) =  -\frac{i }{2} V(\omega) \int_{0}^{t}\mbox{d}t'
\left[b(t') + b^{\dagger}(t')\right]\;e^{-i\omega(t- t')} +
b_{\omega}(0)\;e^{-i\omega t}.
\end{equation}
The annihilation operators are defined in terms of the (transverse) physical
fields:
\begin{eqnarray}\label{physfielddef}
a(t) &  = & \sqrt{\frac{\varepsilon_{0}}{2\hbar \tilde k c}}
\left[\tilde k c A(t) - i E(t)\right], \nonumber \\
b(t) & = & \sqrt{\frac{\rho}{2 \hbar \tilde \omega_{0}}}\left[\tilde
\omega_{0} X(t) + \frac{i}{\rho} P(t)\right],
\end{eqnarray}
and similarly for the creation operators. Here $A$ and $E$ are the vector
potential and the electric field, $X$ the polarization field and $P$ its
canonical conjugate. Insertion of the solution (\ref{bathsolv}) and its
Hermitian conjugate in the equations of motion gives:
\begin{eqnarray}\label{physfieldeq}
\dot E(t) & = &(\tilde k c)^{2} A(t) + (\omega_{c}^{2}/\alpha)
P(t), \nonumber\\
\dot A(t) & = &- E(t), \nonumber \\
\dot X(t) & = &
(\omega_{c}/\alpha)^{2}\varepsilon_{0} P(t) +
             (\omega_{c}^{2}/\alpha)\varepsilon_{0} A(t), \nonumber\\
\dot P(t) & = &
-\alpha^{2}\tilde\omega_{0}^{2}/(\varepsilon_{0}\omega_{c}^{2})
X(t) +\alpha^{2}\tilde\omega_{0}/(2\varepsilon_{0}\omega_{c}^{2})
\int_{0}^{t}\mbox{d}t'\;F(t-t') X(t') -B(t).
\end{eqnarray}
In the last equation, the bath operator $B(t)$ is defined as
\begin{equation}
B(t) \equiv
\sqrt{\frac{\hbar\tilde\omega_{0}\rho}{2}}\int_{0}^{\infty}\mbox{d}\omega_{1}\;
V(\omega_{1})\left[\; b_{\omega_{1}}(0)\;e^{-i\omega_{1}t}
+ b_{\omega_{1}}^{\dagger}(0)\;e^{i\omega_{1}t}\;\right],
\end{equation}
whereas the  the function $F$ in the convolution in (\ref{physfieldeq}) is:
\begin{equation}\label{deff}
F(t) \equiv
2\int_{0}^{\infty}\mbox{d}\omega_{1}\;V^{2}(\omega_{1})\sin(\omega_{1}t).
\end{equation}
We get a system of algebraic equations by taking the Laplace transform, which
we denote by a bar:
\begin{equation}\label{laplacevelden}
\left(
\begin{array}{cccc}
p & -\tilde k^{2}c^{2} & 0 & -\omega_{c}^{2}/\alpha   \\ 1 & p & 0
& 0 \\ 0 & -\varepsilon_{0}\omega_{c}^{2}/\alpha & p &
-\varepsilon_{0}\omega_{c}^{2}/\alpha^{2} \\ 0 & 0 & \alpha^{2}
\tilde \omega_{0}^{2}/\varepsilon_{0}\omega_{c}^{2}\left[ 1- \bar
F(p)/(2\tilde\omega_{0})\right] & p
\end{array}
\right)
\left(
\begin{array}{c}
\bar E(p) \\ \bar A(p) \\ \bar X(p) \\ \bar P(p)
\end{array}
\right)
=
\left(
\begin{array}{c}
E(0) \\ A(0) \\ X(0) \\ P(0) -\bar B(p)
\end{array}
\right).
\end{equation}
Through the operator $\bar B(p)$ the bath remains part of the system of
equations: this is as far as we can ``integrate out'' the bath variables.

Now we can determine the dielectric function $\varepsilon(\omega)$, which is a
{\em classical} quantity, by putting the determinant of the $(4 \times 4)$
coefficient matrix to zero. The determinant gives the dispersion relation
\begin{equation}\label{dispdef}
\bar D(p)\equiv\bar\varepsilon(p)p^{2} +k^{2}c^{2} = 0,
\end{equation}
with the ``Laplace dielectric function''
\begin{equation}\label{epsdef}
\bar\varepsilon(p) = 1 + \frac{\omega_{c}^{2}}{p^{2} +
\tilde\omega_{0}^{2} -\frac{1}{2}\tilde\omega_{0}\bar F(p)}.
\end{equation}
The function $\bar F(p)$ is the Laplace transform of $F(t)$, which was defined
in equation (\ref{deff}). From this we find the dielectric function
\begin{equation}\label{epsilon}
\varepsilon(\omega)= \bar\varepsilon(-i\omega +\eta) = 1 -
\frac{\omega_{c}^{2}}{\omega^{2}-\tilde \omega_{0}^{2}+\frac{1}{2}
\tilde \omega_{0}F(\omega)},
\end{equation}
with infinitesimal positive $\eta$ and
\begin{equation}\label{defF}
F(\omega) \equiv \bar F(-i\omega + \eta)=
\int_{0}^{\infty}\mbox{d}\omega_{1} V^{2}(\omega_{1})\left(
\frac{1}{\omega_{1}-\omega -i\eta} + \frac{1}{\omega_{1}+\omega
+i\eta}\right).
\end{equation}
The difference between  $F(\omega)$ and  $F(t)$ is denoted by their arguments.
The dielectric function  satisfies the Kramers-Kronig-relations and  has the
property of a response function that $\varepsilon(-\omega^{*})$ equals
$\varepsilon^{*}(\omega)$. It can be shown that it has no poles in the upper
half plane, provided that the integral in (\ref{tildeomnul}) exists. Previous
authors \cite{HuttnerB92b,DutraF98,Bechler99} assumed that the analytical
continuation of $V^{2}(\omega)$ to negative frequencies is anti-symmetrical in
frequency. Then (\ref{epsilon}) reduces to the dielectric constant in
\cite{DutraF98}, where it was shown to be identical to the more complicated
expression in \cite{HuttnerB92b}.

We combine (\ref{laplacevelden}) and (\ref{epsdef}) and write the Laplace
fields in terms of the fields at time $t=0$, with coefficients that are
functions of the Laplace dielectric function $\bar\varepsilon(p)$ and
susceptibility $\bar \chi(p)= \bar\varepsilon(p)-1$. For the electric field we
find:
\begin{eqnarray}\label{Einfields0}
\bar E(p) &= & \bar D^{-1}(p) \left\{ \rule{0mm}{5mm} p E(0) +
[p^{2}\bar\chi(p)+k^{2}c^{2}]A(0) \right. \nonumber \\
 & &\left. +\frac{\alpha}{\varepsilon_{0}}p\left[\frac{p^{2}}{\omega_{c}^{2}}
\bar\chi(p)-1\right]X(0)
+ \frac{1}{\alpha}p^{2}\bar\chi(p)[P(0)-\bar B(p)] \right\}.
\end{eqnarray}
The other Laplace operators can be found in the same way and are listed in the
Appendix. The inverse Laplace transform gives the fields at time $t$ in terms
of the fields at time $t=0$:
\begin{equation}\label{timecoef}
E(t) = M_{EE}(t)E(0) + M_{EA}(t)A(0) +M_{EX}(t)X(0) +
M_{EP}(t)P(0) + B_{E}(t),
\end{equation}
where, for instance,
\begin{equation}\label{mee}
M_{EE}(t) =  \frac{1}{2 \pi
i}\int_{-i\infty}^{i\infty}\mbox{d}p\; e^{pt}\; \bar D^{-1}(p) \; p.
\end{equation}
The operator $B_{E}(t)$ in equation (\ref{timecoef}) is the contribution of
the $t=0$ bath operators to the electric field. This term will be analyzed in
more detail in section \ref{longtime}.

The {\em equal} time commutation relations of the field operators
are
\begin{equation}\label{cancom}
[A(\lambda,{\bf k},t), -\varepsilon_{0}E(\lambda', {\bf - k'}, t)]
= [X(\lambda, {\bf k}, t), P(\lambda', {\bf - k'}, t)]= i\hbar
\delta_{\lambda \lambda'}\delta({\bf k}-{\bf k'}).
\end{equation}
All other inequivalent combinations of operators commute. In particular $A$
and $X$ are independent canonical variables. Hence, we have the property $[A,
-D] = [A, -\varepsilon_{0}E]$, with the displacement field $D$ defined as
$\varepsilon_{0}E-\alpha X$. With the help of (\ref{timecoef}) and
(\ref{cancom}), we can also calculate {\em non-equal} time commutators, for
example:
\begin{equation}\label{commeet}
[E(\lambda,{\bf k},t), E(\lambda', {\bf - k'}, 0)] = M_{EA}(t)
[A(\lambda,{\bf k},0), E(\lambda', {\bf - k'}, 0)] =
-\frac{i\hbar}{\varepsilon_{0}}M_{EA}(t)\delta_{\lambda
\lambda'}\delta({\bf k}-{\bf k'}).
\end{equation}

In principle we have solved the complete time evolution of the field
operators. In section \ref{sumrules} we analyze in more detail their
short-time behavior, whereas in section \ref{longtime} we consider the
long-time limit.

\section{Short-time limit:  sum rules}\label{sumrules}
For fixed $k$, the zeroes of the dispersion relation (\ref{dispdef}) are the
poles of the integrand in (\ref{mee}). We assume that they are simple
first-order poles and rewrite the integral (\ref{mee}) as an integral over
frequencies $\omega = i p$. Then, using contour integration in the lower
frequency half plane, we find the coefficients for the electric field:
\begin{eqnarray}\label{timeco2}
M_{EE}(t) & = &
\sum_{j}\mbox{Re}\left[\frac{v_{p,j}v_{g,j}}{c^{2}}
e^{-i\Omega_{j}t}\right], \nonumber\\
 M_{EA}(t) &= &
-kc\sum_{j} \mbox{Im} \left[\frac{v_{p,j}^{2}v_{g,j}}{c^{3}}
e^{-i\Omega_{j}t}\right], \nonumber
\\ M_{EX}(t)& = & -\frac{\alpha
k^{2}c^{2}}{\omega_{c}^{2}\varepsilon_{0}}\sum_{j}\mbox{Re}\;\left[
 \frac{v_{p,j}v_{g,j}}{c^{2}}\left(1-\frac{v_{p,j}^{2}}{c^{2}}
+ \frac{\omega_{c}^{2}}{k^{2}c^{2}}\right)e^{-i\Omega_{j}t}\right],
\nonumber \\
M_{EP}(t)& = & \frac{k c}{\alpha }
\sum_{j}\mbox{Im}\;\left[\frac{v_{g,j}}{c}(1 -
\frac{v_{p,j}^{2}}{c^{2}})e^{-i\Omega_{j}t}\right].
\end{eqnarray}
Some details of the calculation and a list of  coefficients $M_{mn}(t)$ of
other operators can be found in the Appendix. In these expressions, the
frequencies $\Omega_{j}=\Omega_{j}(k)$ are the complex-frequency solutions of
the dispersion relation $\omega^{2}\varepsilon(\omega)-k^{2}c^{2}=0$. All
$\Omega_{j}(k)$ have a negative imaginary part. Since
$\varepsilon(-\omega^{*})=\varepsilon^{*}(\omega)$, it follows that
$-\Omega_{j}^{*}(k)$ is also a solution of the dispersion relation. We can
choose $\Omega_{j}(k)$ to be the solution with a positive real part. The
summation over $j$ is a summation over all the polariton branches of the
medium. For each branch, the complex phase velocity is defined as $v_{p,j}(k)
= \Omega_{j}(k)/k$ and the group velocity as $v_{g,j}(k)
=\mbox{d}\Omega_{j}(k)/\mbox{d}k$. For convenience, we leave out their
explicit $k$-dependence in the following.

\mbox{}From equation (\ref{timecoef}) we can see that the ``diagonal''
coefficient $M_{EE}(t)$ in (\ref{timeco2}) should have the value $1$ at time
$t=0$ and the ``off-diagonal'' coefficients $M_{EA}(0), M_{EX}(0)$, etc.
should have the value $0$. The coefficients of the other field operators
should also follow this rule. If these constraints are satisfied the non-equal
time commutators like (\ref{commeet}) get the right equal-time limits as well.
The coefficients (\ref{timeco2}) can only have the right $t=0$ limits, if
certain velocity sum rules are satisfied.

Velocity sum rules can be derived in a systematic way by evaluating the
following two types of integrals:
\begin{eqnarray}
\int_{-\infty}^{\infty}\mbox{d}\omega\;\frac{(\omega +
i\delta)^{n}}{\varepsilon(\omega)\omega^{2} - k^{2}c^{2}} \qquad &
\mbox{for}\;n=-1,0,1, \label{firsttype}\\
\int_{-\infty}^{\infty}\mbox{d}\omega\;\frac{(\omega +
i\delta)^{m}\chi(\omega)}{\varepsilon(\omega)\omega^{2} -
k^{2}c^{2}} \qquad & \mbox{for}\;m=-1,0,1,2,3.\label{secondtype}
\end{eqnarray}
Here $\varepsilon(\omega)$ is an {\em arbitrary} dielectric function that
satisfies the Kramers-Kronig relations, so it is not necessarily of the
specific form (\ref{epsilon}). The integrals can be evaluated  using contour
integration in the complex frequency plane. We can close the contours either
in the upper or in the lower half plane. Equating the two answers gives a
velocity sum rule. In this way one finds for all wavevectors $k$:
\begin{eqnarray}\label{HBsumrules}
\sum_{j}\mbox{Re}(v_{g,j}v_{p,j}/c^{2}) = 1,  \label{HBsum1}\\
\sum_{j}\mbox{Re}(v_{g,j}/v_{p,j}) = 1.  \label{HBsum2}
\end{eqnarray}
These sum rules can be found from (\ref{firsttype}) with $n=1$ and $n=-1$,
respectively. Both relations have been obtained before
\cite{HuttnerBB91,HuttnerB92b,AlDossaryBE96,DrummondH99}. The second was
coined the Huttner-Barnett sum rule in \cite{AlDossaryBE96}, because of its
importance in phenomenological quantum theories of dielectrics. A second group
of sum rules has the form
\begin{equation} \label{imrule1}
 \sum_{j}\mbox{Im}(v_{g,j}v^{2 q}_{p,j}) = 0\;\qquad\forall\,
q=-1, 0,1.
\end{equation}
The rules with $q=-1$ and $q=1$ follow from (\ref{secondtype}) with $m=0$ and
$m=2$, respectively; the case with $q=0$ follows from (\ref{firsttype}) with
$n=0$.

All of these sum rules  are independent of any specific form of the dielectric
function, as long as it satisfies the Kramers-Kronig relations. Other sum
rules do depend on the behavior of $\varepsilon(\omega)$ for high or low
frequencies. For example, from (\ref{secondtype}) with $m=-1$ we find:
\begin{equation}\label{epsnulrule}
\sum_{j}\mbox{Re}(c^2 v_{g,j}/v_{p,j}^{3}) = \varepsilon(0).
\end{equation}
This sum rule  depends on the static limit of the dielectric function. For
conductors the dielectric function is singular at $\omega=0$ \cite{Jackson75},
but for dielectric functions which can be found from the damped-polariton
model, $\varepsilon(0)$ is finite. Two other sum rules can be derived when for
high frequencies $\omega^{2}\chi(\omega)$ approaches a constant value that we
name $-\omega_{\rm lim}^{2}$. From (\ref{secondtype}) with $m=3$ we then find
\begin{equation} \label{epsinfrule}
\sum_{j}\mbox{Re}(v_{g,j}v^{3}_{p,j}/c^{4}) = 1 +
(\omega_{\rm lim}/k c)^{2}.
\end{equation}
Moreover, if $\omega^{2}\chi(\omega)+\omega_{\rm lim}^{2}$ falls off faster
than $\omega^{-1}$, then  the integral
\begin{equation}\label{thirdtype}
\int_{-\infty}^{\infty}\mbox{d}\omega\;\frac{\omega^{2}[\omega^{2}\chi(\omega)+
\omega_{\rm lim}^{2}]}
{\varepsilon(\omega)\omega^{2}- k^{2}c^{2}}
\end{equation}
produces the sum rule (\ref{imrule1}) with $q=2$.

Returning now to the time-dependent coefficients (\ref{timeco2}) (and the
other ones in the Appendix), one finds by inspection that one needs all the
above sum rules except (\ref{epsnulrule}) to prove that the coefficients have
the right limits for $t=0$. In particular, from equation (\ref{epsilon}) it
follows that the frequency $\omega_{\rm lim}$ as defined above exists in the
damped-polariton model and equals $\omega_{c}$. Then with (\ref{HBsum2}) and
(\ref{epsinfrule}) we see that indeed one has $M_{EX}(0)=0$ in
(\ref{timeco2}).

It is easy to prove the above sum rules in  the following one-resonance model:
\begin{equation}\label{realeps}
\varepsilon(\omega) = 1 - \frac{\omega_{c}^{2}}{\omega^{2} -
\omega_{0}^{2}}.
\end{equation}
This $\varepsilon(\omega)$ is real and violates the Kramers-Kronig relations,
but it can be considered as a limiting case of an acceptable dielectric
function. The high-frequency limit of $\omega^{2}\chi(\omega)$ indeed equals
$-\omega_{c}^{2}$. The two sum rules (\ref{HBsum1}), (\ref{HBsum2}) were shown
to be valid for this model \cite{HuttnerB92b} and we want to check
(\ref{epsinfrule}) as well. The dispersion relation is
\begin{equation}\label{dispsimp}
\omega^{4} -(\omega_{0}^{2} + \omega_{c}^{2}
+k^{2}c^{2})\omega^{2} + k^{2}c^{2}\omega_{0}^{2} = 0,
\end{equation}
which has two (real) solutions $\Omega_{+}^{2}$ and $\Omega_{-}^{2}$ with sum
$(\omega_{0}^{2} + \omega_{c}^{2} +k^{2}c^{2})$ and product
$k^{2}c^{2}\omega_{0}^{2}$. It follows that for all $k$
\begin{equation}\label{rulecheck}
v_{p,+}^{3}v_{g,+}+ v_{p,-}^{3}v_{g,-} =
\frac{1}{4k^{3}}\frac{\mbox{d}}{\mbox{d}k}(\Omega_{+}^{4}+\Omega_{-}^{4})
=
\frac{1}{4k^{3}}\frac{\mbox{d}}{\mbox{d}k}\left[(\Omega_{+}^{2}+\Omega_{-}^{2})^{2}
-2 \Omega_{+}^{2}\Omega_{-}^{2}\right] = \left(1 +
\frac{\omega_{c}^{2}}{k^{2}c^{2}}\right)c^{4},
\end{equation}
in agreement with (\ref{epsinfrule}). The other sum rules can also be checked
for this simple model. The sum rules (\ref{imrule1}) obviously hold, because
all group and phase velocities are real in this model. In models that respect
the Kramers-Kronig-relations, these sum rules are nontrivial.

\section{Long-time limit}\label{longtime}
\subsection{Field and medium operators}\label{fieldslong}

The coefficients $M_{EE}(t)$ etc. in (\ref{timeco2})  damp out exponentially
in time. Every polariton branch has its own characteristic damping time
$\tau_{j}(k)=1/(\mbox{Im}\;\Omega_{j}(k))$. After a few times the maximum
characteristic damping period, with the maximum taken over all branches, the
exponentially damped coefficients can be neglected. We call this the long-time
limit. The speed at which it is attained, depends on $\varepsilon(\omega)$ and
on $k$. For long times, {\em only} the bath operator $B_{E}(t)$  in
(\ref{timecoef}) survives, because it has poles on the imaginary axis in the
complex $p$-plane:
\begin{equation}\label{badpolen}
B_{E}(t) = -\frac{1}{2\pi i\omega_c}
\sqrt{\frac{\hbar\tilde\omega_{0}}{2\varepsilon_0}}
\int_{0}^{\infty}\mbox{d}\omega_{1}\;V(\omega_{1})
\int_{-i\infty}^{i\infty}\mbox{d}p\;e^{pt}\;
\frac{p^{2}\bar\chi(p)}{\bar\varepsilon(p)p^{2}+k^{2}c^{2}}
\left[\frac{b_{\omega_{1}}(0)}{p+i\omega_{1}}
+\frac{b_{\omega_{1}}^{\dagger}(0)}{p-i\omega_{1}} \right].
\end{equation}
Hence, in the long-time limit, all field operators are functions of the
initial bath operators alone. For the electric field we find
\begin{equation}\label{Elongtime}
E(t) \rightarrow E_{l}(t) = -\frac{1}{\omega_{c}}
\sqrt\frac{\hbar\tilde \omega_{0}}{2 \varepsilon_{0}}
\int_{0}^{\infty}\mbox{d}\omega_{1}\;V(\omega_{1}) \left[
\frac{\omega_{1}^{2}\chi(\omega_{1})b_{\omega_{1}}(0)e^{-i\omega_{1}t}}
{\varepsilon(\omega_{1})\omega_{1}^{2} - k^{2}c^{2}} +
\frac{\omega_{1}^{2}\chi^{*}(\omega_{1})b_{\omega_{1}}^{\dagger}(0)e^{i\omega_{1}t}}
{\varepsilon^{*}(\omega_{1})\omega_{1}^{2} - k^{2}c^{2}}\right],
\end{equation}
where the subscript $l$ denotes the long-time limit. The temporal (and
spatial)  Fourier components of the long-time solutions are:
\begin{eqnarray}\label{longtimefourier}
E_{l}^{+}(\omega) & = & -\frac{1}{\omega_{c}}
\sqrt{\frac{\hbar\tilde \omega_{0}}{2 \varepsilon_{0}}}
\frac{V(\omega)\omega^{2}\chi(\omega)b_{\omega}(0)}
{\varepsilon(\omega)\omega^{2} - k^{2}c^{2}}, \nonumber \\
A_{l}^{+}(\omega) & = & \frac{i}{\omega_{c}}
\sqrt{\frac{\hbar\tilde \omega_{0}}{2 \varepsilon_{0}}} \frac{
V(\omega)\omega\chi(\omega)b_{\omega}(0)}
{\varepsilon(\omega)\omega^{2} - k^{2}c^{2}}, \nonumber \\
X_{l}^{+}( \omega ) & = & -\frac{1}{\alpha\omega_{c}}
\sqrt{\frac{\hbar\tilde \omega_{0} \varepsilon_{0}}{2}} \frac{
V(\omega)(\omega^{2}-k^{2}c^{2})\chi(\omega)b_{\omega}(0)}
{\varepsilon(\omega)\omega^{2} - k^{2}c^{2}}, \nonumber \\
P_{l}^{+}(\omega) & = &\frac{i\alpha}{\omega_{c}^{3}}
\sqrt{\frac{\hbar\tilde \omega_{0} }{2\varepsilon_{0}}} \frac{
V(\omega)(\omega^{2}-k^{2}c^{2}-\omega_{c}^{2})\omega\chi(\omega)b_{\omega}(0)}
{\varepsilon(\omega)\omega^{2} - k^{2}c^{2}},
\end{eqnarray}
where the superscript $+$ denotes the positive-frequency component of the
operator. For future reference we also give the long-time limit of the
electric field operator as a function of position and time:
\begin{eqnarray}\label{fieldlongtime}
\lefteqn{{\bf E}_{l}({\bf r}, t) =
-\sqrt{\frac{\hbar\tilde\omega_{0}}{2(2\pi)^{3}\varepsilon_{0}\omega_{c}^{2}}}
}\nonumber\\
&&
\times\int\mbox{d}{\bf k}\sum_{\lambda =1,2}{\bf e}_{\lambda}({\bf k})
\int_{0}^{\infty}\mbox{d}\omega_{1} \left[
\frac{V(\omega_{1})\omega_{1}^{2}\chi(\omega_{1})b_{\omega_{1}}(\lambda,{\bf
k}, 0)}{\varepsilon(\omega_{1})\omega_{1}^{2}-k^{2}c^{2}}
e^{i({\bf k\cdot r}-\omega_{1}t)} + \mbox{H.c.}\;\right].\rule{8mm}{0mm}
\end{eqnarray}
Similar expressions can be given for the other operators. Notice that these
long-time solutions indeed are solutions of the equations of motion
(\ref{physfieldeq}) and of the Maxwell equations. The canonical commutation
relations (\ref{cancom}) should be preserved in this long-time limit. Also,
the  non-equal time commutation relations like in equation (\ref{commeet})
should be time-translation invariant.  The commutation relations can be
verified with the equality
\begin{equation}\label{subsv}
\frac{\pi\tilde\omega_{0}}{2
\omega_{c}^{2}}V^{2}(\omega)\left|\chi(\omega)\right|^{2} =
 \mbox{Im}\; \chi(\omega) = \mbox{Im}\;
\varepsilon(\omega) \equiv \varepsilon_{i}(\omega),
\end{equation}
which follows from equations (\ref{epsilon}) and (\ref{defF}). Since
$\varepsilon_{i}(\omega)$ is anti-symmetric in $\omega$, all commutators can
be shown to be proportional to integrals over the whole real frequency axis.
Contour integration then leads to the required results.

The solutions found above can be related to those obtained by explicit
diagonalization of the full Hamiltonian of the model. In \cite{HuttnerB92b}
this diagonalization was carried out by using Fano's technique. In that way
the field and medium operators were written in terms of the diagonalizing
annihilation operators (called  $C({\bf k},\omega)$ in \cite{HuttnerB92b}) and
the corresponding creation operators. If one replaces the bath annihilation
operators  $b_{\omega}({\bf k}, 0)$ in the long-time  solutions
(\ref{longtimefourier}) by the diagonalizing annihilation operators $C({\bf
k},\omega)$, and if one makes similar replacements for the creation operators,
the expressions for the field and medium operators in \cite{HuttnerB92b} are
recovered.

The long-time solutions can be interpreted as follows: when the dielectric
medium is prepared in a state that is not an eigenstate of the Hamiltonian and
if the coupling $V(\omega)$ is nonzero for all frequencies, then the medium
tends to an equilibrium that is determined by the state of the bath. The time
it takes for this equilibrium to settle down is the time after which the
long-time solutions can be used for the field operators. So one can always use
the long-time solutions in the calculations, unless the medium has been
specially prepared in a non-equilibrium state a short time before one does the
experiment. The interpretation of the long-time solution will become clearer
in section \ref{SE} where we calculate spontaneous emission.

In summary, for times long after $t=0$, all field operators can be expressed
solely in terms of the bath operators at time $t=0$. The time evolution is
governed by the bath Hamiltonian alone. The field operators still satisfy
Maxwell's equations and the canonical commutation relations. Classical
expressions for the Maxwell fields  would have died exponentially to zero in
this long time limit.

\subsection{Relation with phenomenological theories}\label{relphen}

The long-time solutions of the field operators can be related to expressions
in phenomenological theories, as we will show presently. In phenomenological
quantum mechanical theories of homogeneous absorbing dielectrics
\cite{GrunerW95} -- \cite{GrunerW96}, a noise current density operator ${\bf
J}$ is added to the Maxwell equations in order to preserve the field
commutation relations:
\begin{eqnarray}\label{Maxphen}
\nabla \times {\bf E} ^{+}({\bf r}, \omega) & = & i \omega {\bf
B}^{+}({\bf r}, \omega), \\
\nabla\times {\bf B}^{+}({\bf r},\omega)
& = & -i\omega\mu_{0}\tilde {\bf D}^{+}({\bf r}, \omega) +
\mu_{0}{\bf J}^{+}({\bf r}, \omega).
\end{eqnarray}
The displacement field $\tilde {\bf D}^{+}$ in the last equation is defined in
terms of the electric field and the dielectric function as
\begin{equation}\label{DEphen}
\tilde {\bf D}^{+}({\bf r}, \omega) =
\varepsilon_{0}\varepsilon(\omega){\bf E}^{+}({\bf r}, \omega).
\end{equation}
We write  ${\bf \tilde D}$ to stress the difference with the microscopic
displacement field ${\bf D}$ in section \ref{model}. After taking the spatial
Fourier transform, and using ${\bf B}^{+} = \nabla \times {\bf A}^{+}$ and
${\bf E}^{+} = i\omega {\bf A}^{+}$, so that the first of the Maxwell
equations is satisfied, one finds from the second equation:
\begin{equation}\label{aphen}
\left[\;\omega^{2}\varepsilon(\omega) -
k^{2}c^{2}\;\right]\;A^{+}(\lambda,{\bf k}, \omega) =
-\frac{1}{\varepsilon_{0}}J^{+}(\lambda, {\bf k}, \omega).
\end{equation}
The vector potential and all Maxwell fields can be calculated in terms of the
noise current density $J$. The canonical commutation relations are preserved,
if for the noise current one chooses \cite{GrunerW95,MatloobL95}:
\begin{equation}\label{jjcomm}
\left[\;J^{+}(\lambda, {\bf k},\omega),[J^{+}(\lambda',{\bf
k'},\omega')]^{\dagger}\;\right]
=
\frac{\hbar\omega^{2}\varepsilon_{0}\varepsilon_{i}(\omega)}{\pi}\;
\delta_{\lambda\lambda'}\delta({\bf k} - {\bf
k'})\delta(\omega-\omega').
\end{equation}
Instead of using the noise current operator, one defines basic bosonic
operators
\begin{equation}\label{fko}
f(\lambda, {\bf k},\omega) = \sqrt{\frac{\pi}{\hbar
\omega^{2}\varepsilon_{0}\varepsilon_{i}(\omega)}}\;J^{+}(\lambda,
{\bf k}, \omega),
\end{equation}
so that these operators satisfy simple commutation relations:
\begin{equation}
[\;f(\lambda, {\bf k},\omega),f^{\dagger}(\lambda',{\bf
k'},\omega')\;] = \delta_{\lambda\lambda'}\delta({\bf k} - {\bf
k'})\delta(\omega-\omega').
\end{equation}

Now we turn to the long-time solutions of the field operators that we
determined in section \ref{fieldslong}. The long-time solution of the vector
potential in (\ref{longtimefourier}) obviously is a solution of the
following inhomogeneous wave equation:
\begin{equation}\label{awaveeq}
\left[\;\varepsilon(\omega) \omega^{2} -
k^{2}c^{2}\;\right]\;A_{l}^{ +}(\lambda, {\bf k},\omega) =
\frac{i}{\omega_{c}}\sqrt{\frac{\hbar\tilde\omega_{0}}{2\varepsilon_{0}}}\;
V(\omega)\omega\chi(\omega)b_{\omega}(\lambda, {\bf k},0).
\end{equation}
This kind of equation is well-known in Langevin theories
\cite{Gardiner91,MandelW95}: the coupling to a bath gives a damping term
(here: a complex dielectric constant) in the equations of motion of the
system. Besides damping, there is an extra term that is neglected classically.
This term is the quantum noise operator, which features the bath operators at
time $t=0$.

The long-time solution (\ref{awaveeq}) can justify the phenomenological
equation (\ref{aphen}), if we identify
\begin{equation}\label{equalnoise}
f_{l}(\lambda, {\bf k}, \omega) = -i
\frac{V(\omega)\chi(\omega)}{|V(\omega)\chi(\omega)|}
b_{\omega}(\lambda, {\bf k}, 0),
\end{equation}
where we used equation (\ref{subsv}). We see that up to a phase factor, the
bath operators $b_{\omega}(\lambda, {\bf k}, 0)$ from the microscopic theory
serve as basic bosonic operators $f(\lambda, {\bf k}, \omega)$ in the
phenomenological theories. We want to stress that the identification
(\ref{equalnoise}) is only valid in the long-time limit when the medium is in
equilibrium with the bath.

In section \ref{model} we saw that $-\varepsilon_{0}E$ is the canonical
conjugate field of $A$ and that $[A, -D]$ gives the canonical result as well.
Since we can make the identification (\ref{equalnoise}), the same relations
hold in the phenomenological theory that was described in this section. But
now let us calculate the commutator $[A, -\tilde D]$ with $\tilde D^{+}$
defined as in equation (\ref{DEphen}) and $\tilde D^{-}$ as its Hermitian
conjugate. We can use the long-time solutions, because the commutation
relations are preserved:
\begin{equation}\label{commad2}
\left[ A(\lambda, {\bf k},t), -\tilde D(\lambda', -{\bf k'},
t)\right] = \frac{2i\hbar}{\pi}\delta_{\lambda\lambda'}\delta({\bf
k}-{\bf k'})
\int_{0}^{\infty}\mbox{d}\omega\;\frac{\varepsilon_{r}(\omega)
\varepsilon_i(\omega)\omega^{3}}
{|\varepsilon(\omega)\omega^{2}-k^{2}c^{2}|^{2}},
\end{equation}
with $\epsilon_r(\omega)$ the real part of the dielectric constant. The
symmetry of the integrand enables us to rewrite the right-hand side as an
integral over all real frequencies. When using contour integration, one cannot
replace $\varepsilon^{*}(\omega)$ by $\varepsilon(-\omega^{*})$, but the
analytical continuation to complex frequencies of
$\varepsilon^{*}(\omega)=\varepsilon(-\omega)$ must be used instead:
\begin{equation}\label{commad3}
\left[ A(\lambda, {\bf k},t), -\tilde D(\lambda', -{\bf k'},
t)\right] = i\hbar \delta_{\lambda\lambda'}\delta({\bf k}-{\bf
k'}) \sum_{j}
\mbox{Re}\;\left[\varepsilon(-\Omega_{j})v_{p,j}v_{g,j}/c^{2}\right],
\end{equation}
where we assumed as before that all poles of the dispersion relation are
first-order poles. Note that $\varepsilon(-\Omega_{j})$ depends on the
behavior of the dielectric function in the {\em upper} half plane. Contrary to
a statement in \cite{MatloobL95}, the commutator does not give the canonical
result, because in general there is no sum rule for the right-hand side of the
equation. In other words, $(D-\varepsilon_{0}E)$ is canonically independent
from $E$, but $(\tilde D - \varepsilon_{0}E)$ is not. The operator $(D -
\tilde D)$ is proportional to the Langevin noise term in the wave equation for
the electric field.

Now let us neglect absorption at all frequencies. Strictly speaking, the limit
$\varepsilon_{i}(\omega)\rightarrow 0$ is unphysical because it violates the
Kramers-Kronig relations, but the limit is sometimes taken for dielectrics
that show negligible absorption at optical frequencies
\cite{GrunerW96,AlDossaryBE96}. When $\varepsilon(\omega)$  becomes
real, the solutions $\Omega_{j}$ become real and in that limit one has
$\varepsilon(-\Omega_{j})\rightarrow \varepsilon(\Omega_{j}) = (c/v_{p,
j})^{2}$. Inserting this in (\ref{commad3}) and using the Huttner-Barnett sum
rule $\sum_{j}\mbox{Re}\;(v_{g,j}/v_{p,j}) = 1$, we immediately find the
canonical result for  $[A, -\tilde D]$. We compare this with the results in
\cite{AlDossaryBE96}, where the dielectric function is assumed to be real.
There a phenomenological Lagrangian was introduced and the  fields $A$ and
$-\tilde D$ were correctly identified as a canonical pair. The Huttner-Barnett
sum rule was invoked to show that their commutator indeed had the canonical
form. It was concluded that it is misleading that also $[A,
-\varepsilon_{0}E]$ has the canonical form. Here we have learnt that this
misleading result is not surprising: in the limit of real dielectric constants
{\em and only then}, both $[A,-\varepsilon_{0}E]$ and $[A,-\tilde D]$ can have
the canonical form in the same gauge, the reason being that $\tilde D$
approaches $D$ in that limit.

\section{Model dielectric functions}\label{modeleps}

Phenomenological theories as discussed in section \ref{relphen} have
expressions for $\varepsilon(\omega)$ as input. In practice, this input will
be the outcome of measurements of the dielectric function. By choosing the
appropriate microscopic coupling constants and resonance frequencies in the
damped-polariton model, one can hope to find a given dielectric function, thus
providing a connection with phenomenological theories. It was argued in
\cite{DutraF98} that the well-known Lorentz oscillator form of the dielectric
function could not be found from the damped-polariton theory in this way. We
shall reconsider this issue below.

A dielectric function that follows from the  damped-polariton Hamiltonian
(\ref{HHHH}) will have a single resonance, because there is only one resonance
frequency $\omega_{0}$ in the matter fields. Experimentally, one may find more
resonances in the $\varepsilon(\omega)$. This should not be used as an
objection to the damped-polariton model, because in principle one could easily
extend the theory with more material resonances. In this section, we consider
two of these one-resonance models.

\subsection{The Lorentz oscillator model}\label{loreps}

We want to find microscopic coupling constants in the damped-polariton theory
so that the resulting $\varepsilon(\omega)$ has the following Lorentz
oscillator form:
\begin{equation}\label{epslor}
\varepsilon_{\rm Lor}(\omega) = 1 -
\frac{\omega_{c,{\rm Lor}}^{2}}{\omega^{2}-\omega_{\rm res}^{2}
+2i\omega \kappa_{0}}.
\end{equation}
Here $\omega_{\rm res}$ is the  resonance frequency of the medium and
$\omega_{c,{\rm Lor}}$ is a frequency that is related to the coupling strength
between the electromagnetic and the matter field. Identifying
$\varepsilon(\omega)$ from equation (\ref{epsilon}) with $\varepsilon_{\rm
Lor}(\omega)$, we find apart from the trivial identification $\omega_{c} =
\omega_{c,{\rm Lor}}$
\begin{equation}\label{HBisL}
\int_{0}^{\infty}\mbox{d}\omega_{1} V^{2}(\omega_{1}) \left[
\frac{1}{\omega_{1}-\omega-i\eta} + \frac{1}{\omega_{1} +\omega
+i\eta}\right] = 4\left(\frac{\omega}{\tilde
\omega_{0}}\right)i\kappa_{0} +\Delta,
\end{equation}
where the frequency shift $\Delta$ is defined such that $\omega_{\rm res}^{2}
= \tilde \omega_{0}^{2} - \tilde \omega_{0}\Delta/2$. The coupling
$V^{2}(\omega_{1})$ is fixed by the identification of the imaginary parts and
for all frequencies it equals $V^{2}(\omega_{1}) = 4\kappa_{0}\omega_{1}/(\pi
\tilde \omega_{0})$. However, if we insert this coupling in the equation for
the real parts, we find that the frequency shift $\Delta$ is infinitely large.
Also, the renormalized frequency $\tilde\omega_{0}$ in equation
(\ref{tildeomnul}) blows up. We can solve this problem  by introducing a
frequency cut-off in the coupling, namely $V^{2}(\omega_{1}) =
4\kappa(\omega_{1})\omega_{1}/(\pi \tilde \omega_{0})$ with
\begin{equation}\label{kappareno}
\kappa(\omega_{1}) =
\frac{\kappa_{0}\Omega^{2}}{\Omega^{2}+\omega_{1}^{2}}.
\end{equation}
With this choice one finds $\tilde\omega_{0}=\sqrt{\omega_{0}^{2}
+2\kappa_{0}\Omega}$, which clearly has a strong  dependence on the cut-off
frequency. The shift $\Delta$ becomes both finite and frequency-dependent:
\begin{equation}
\Delta(\omega) = \frac{4}{\pi\tilde\omega_{0}}\;  {\cal P}
\int_{0}^{\infty}\mbox{d}\omega_{1}\;\omega_{1}\kappa(\omega_{1})
\left(\frac{1}{\omega_{1}-\omega}
+\frac{1}{\omega_{1}+\omega}\right) =
4\kappa(\omega)\frac{\Omega}{\tilde\omega_{0}}. \end{equation}
The principal value integral can be evaluated by means of contour integration
in the complex frequency plane. In this way we arrive at the following
expression for the dielectric function:
\begin{equation}\label{sumepslor}
\varepsilon(\omega) = 1 - \frac{\omega_{c}^{2}} {\omega^{2}
-\omega_{0}^{2}-2\Omega[\kappa_{0} - \kappa(\omega)] + 2 i
\omega\kappa(\omega)}.
\end{equation}
We can choose $\Omega$ arbitrarily high (but finite). Quite unlike $\tilde
\omega_{0}$,  the optical resonance frequency $\omega_{\rm res}$ approaches
$\omega_{0}$ from above, the higher we choose the cut-off, since one has:
$\omega_{\rm res}^{2} \simeq \omega_{0}^{2} +
2\kappa_{0}\omega_{0}^{2}/\Omega$. Note that the dielectric function
(\ref{sumepslor}) has the right high-frequency limit $\omega^{2}\chi(\omega)
\rightarrow -\omega_{c}^{2}$ as required in section \ref{sumrules}.

It is well-known that there are two branches of solutions of the
dispersion relation when the dielectric function is of the form
(\ref{epslor}): there is an upper and a lower polariton branch.
The dielectric function (\ref{sumepslor}) gives rise to {\em
another} branch: it has a purely imaginary frequency with
magnitude of the order of the cut-off frequency. This ``cut-off
branch'' has negligible $k$-dependence. In fact, the leading
$k$-dependent term for large $\Omega$ is $2i\omega_c^2\kappa_0
k^2c^2/\Omega^4$. Clearly, the group velocity on this branch is
practically zero, so that the contribution of the cut-off branch
to the velocity sum rules of section \ref{sumrules} can be
neglected.

We conclude that high cut-off frequencies can be chosen such that
in the optical frequency regime the dielectric function cannot be
discerned from a Lorentz dielectric function with resonance
frequency $\omega_{\rm res}= \omega_{0}$ and damping constant
$\kappa_{0}$. The solutions of the dispersion relation of the
upper and the lower polariton branch together satisfy the sum
rules of section \ref{sumrules}.

\subsection{The point scattering model}\label{emeps}

In general the dielectric function $\varepsilon(\omega)$ describes the
propagation of a coherent light beam in a fixed direction in an isotropic
medium. A complex $\varepsilon(\omega)$ means that there is extinction, which
can be caused either by scattering or absorption, or both. The dielectric
function does not contain information about the extinction mechanism. A
well-known dielectric medium showing polariton behavior is the dilute gas,
which can be described as a collection of point dipoles that scatter light
independently. If only one type of elastic scatterers is present, each having
only one resonance, then the dielectric function is given by
\cite{LagendijkT96}:
\begin{equation}\label{epsisa}
\varepsilon_{{\rm sc}}(\omega) = 1 - \frac{4\pi c^{2}\Gamma_e n}
{\omega^{2} - \omega_{\rm res}^{2} + \frac{2}{3}i\Gamma_e
\omega^{3}/c},
\end{equation}
where $n= N/V$ is the density of the scatterers (not to be
confused with the refractive index $n(\omega)$) and $\Gamma_e =
e^{2}/(4\pi\varepsilon_{0}m_{e}c^{2})$ is the classical electron
radius. This dielectric function can also be found if one supposes
that the medium consists of classical harmonically bound point
charges whose motion is described by the Abraham-Lorentz equation.
The dielectric function (\ref{epsisa}) has the property that the
corresponding $T$-matrix $t(\omega)$ satisfies the optical
theorem, with $t(\omega)$ defined as $\varepsilon(\omega) = 1 - n
t(\omega)(c/\omega)^{2}$. However, (\ref{epsisa}) is not a proper
response function, since it has a pole near the very large {\em
positive} imaginary frequency $3 i c/(2 \Gamma_e)$. This can be
related to the need for the a-causal phenomenon called
pre-acceleration to avoid so-called runaway solutions of the
Abraham-Lorentz equation \cite{Rohrlich65}.

Although we know that in the damped-polariton theory only proper
response functions can be found, we proceed like in the previous
subsection and try to find coupling constants that in the optical
regime give rise to the dielectric function (\ref{epsisa}).
Equating with (\ref{epsilon}) we get $\omega_{c}^{2} = 4\pi c^2
\Gamma_e n$ and $V^{2}(\omega_{1}) =4
\Gamma(\omega_1)\omega_1^3/(3\pi \tilde\omega_{0}c)$, with
\begin{equation}\label{visa}
\Gamma(\omega_1)=\frac{\Gamma_e\Omega^4}{\Omega^4+\omega_1^4}.
\end{equation}
Here we have inserted a convenient frequency cut-off from the
start in order to keep finite the frequency $\tilde \omega_{0}$
and the shift $\Delta$. Contour integration gives
$\tilde\omega_{0}^{2} = \omega_{0}^{2}
+\sqrt{2}\Gamma_e\Omega^{3}/(3c)$, and
\begin{equation}\label{deltaisa}
\Delta(\omega) = \frac{2\sqrt{2}}{3\tilde\omega_{0}
c}\Gamma(\omega)\Omega(\Omega^2+\omega^2).
\end{equation}
The dielectric function has the form
\begin{equation}\label{epsem}
\varepsilon(\omega) = 1 - \frac{4\pi c^{2} \Gamma_e n}
{\omega^{2}- \omega_{0}^{2}
+[\sqrt{2}\Omega^{3}(\Omega^2-\omega^2)/(3\omega^2
c)][\Gamma_e-\Gamma(\omega)]+\frac{2}{3}i\Gamma(\omega)
\omega^{3}/c}.
\end{equation}
In this case, the resonance frequency shifts to frequencies lower
than $\omega_{0}$ and the shift is larger for larger cut-off
frequencies. However, since the classical electron radius is so
much smaller than an optical wavelength, it is very well possible
to choose a cut-off frequency such that $\omega_{0} \ll \Omega \ll
c/\Gamma_e$. Then for optical frequencies, the dielectric function
(\ref{epsem}) is of the form (\ref{epsisa}). Note that for high
frequencies  $\omega^{2}\chi(\omega) \rightarrow -\omega_{c}^{2}$
for the dielectric function (\ref{epsem}), but not for
(\ref{epsisa}).

\begin{figure}
\begin{center}
{\includegraphics[width=85mm, height=60mm]{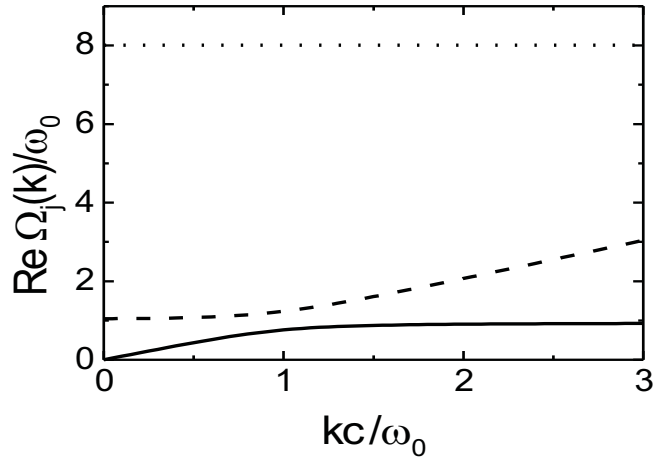}}
\end{center}
\caption{\small Real parts of the three solutions $\Omega_{j}$ of
the dispersion relation with $\varepsilon(\omega)$ as in
(\ref{epsem}).  Numerical values of the parameters: $\Omega =
10\omega_{0}$, $\omega_{c}= 0.5\omega_{0}$ and $\kappa = 0.01
\omega_{0}$. The solid line is the lower polariton branch, the
upper polariton branch is dashed and the cut-off branch is dotted.
}\label{fig1}
\end{figure}

\begin{figure}
\begin{center}
{\includegraphics[width=85mm, height=60mm]{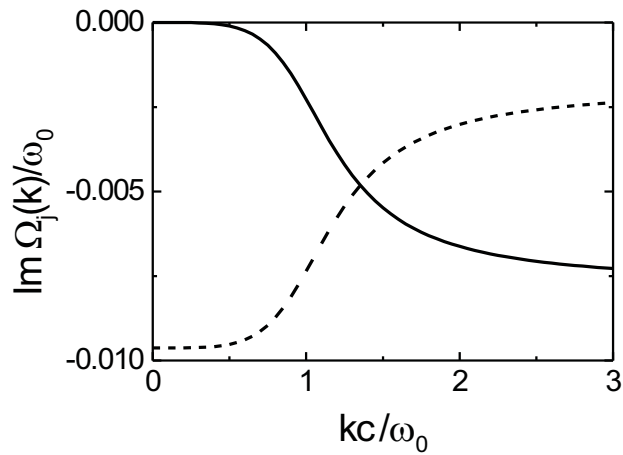}}
\end{center}
\caption{\small Imaginary parts of the lower (solid line) and
upper (dashed line) polariton solutions $\Omega_{j}$ of the
dispersion relation with $\varepsilon(\omega)$ as in
(\ref{epsem}). Numerical values of the parameters as in the
previous plot. Not shown is the imaginary part of the cut-off
branch,  which is also negative and about a thousand times larger
in magnitude. }\label{fig2}
\end{figure}

Again, the frequency cut-off introduces a cut-off branch. In Fig.\
\ref{fig1} we plot the real parts of the three solutions
$\Omega_{j}(k)$ of the dispersion relation. As a measure of the
damping, we introduce $\kappa$ which is given by
$\Gamma_e\omega_{0}^{2}/(3 c)$. For purpose of presentation, the
numerical values of both $\omega_{c}$ and $\kappa$ were chosen
artificially large for a dilute gas. The frequencies on the
cut-off branch are of the same magnitude as the cut-off frequency
$\Omega$, much higher than the optical regime. The imaginary parts
of the upper and lower polariton branches are plotted in Fig.\
\ref{fig2}. The imaginary part of the cut-off branch is large
negative and practically constant  for parameters as given in
Fig.\ \ref{fig1}. Again, since the group velocity on the cut-off
branch is practically zero, the upper and lower polariton branches
together satisfy the sum rules of section \ref{sumrules}. In
particular, Fig.\ \ref{fig2} illustrates that the upper and lower
polariton group velocities $v_{g, u}$ and $v_{g, l}$ satisfy the
sum rule $\mbox{Im }(v_{g, u} + v_{g,l})=0$. 

The cut-off, which was necessary to produce the dielectric function in the
damped-polariton theory, neatly removes the pre-acceleration behavior
associated with a pole in the upper halfplane and leads to a good response
function. The form of the coupling $V(\omega)$ given above (\ref{visa}) has
the following physical interpretation. By equating the damped-polariton
dielectric function with (\ref{epsisa}), we assumed that the dilute gas can be
described as a homogeneous dielectric. The light scattering by the gas
molecules can be accounted for by an absorptive coupling to the {\em free}
electromagnetic field, as long as only single scattering of light is relevant.
Then scattered light is lost for propagation in the original direction. If the
matter-bath coupling is dipole coupling, then for optical frequencies the
product $V^{2}(\omega_{1})/\omega_{1}$ should be proportional to the density
of states of the electromagnetic field, which goes quadratically in frequency.
This is indeed the case.

\section{Spontaneous emission}\label{SE}

The spontaneous emission rate in principle is a time-dependent quantity. In
this section we investigate the transient dynamics of the spontaneous emission
rate of a guest atom in an absorbing medium, when the transition frequency of
the guest atom is close to a material resonance of the medium. We show how our
results relate to previous treatments of spontaneous emission in absorbing
dielectrics, where Fermi's Golden Rule was used to show that the
time-independent (equilibrium) value for the spontaneous emission rate equals
$\Gamma_{0}\mbox{Re}[n(\omega_{A})]$ \cite{BarnettHLM96,BarnettHL92}.
Recently, local field effects have been included in quantum electrodynamical
formulations of the problem
\cite{YablonovitchGB88,KnoesterM89,Milonni95,BarnettHLM96,Fleischhauer99},
but we shall not focus on them in this paper.

We model the guest atom as a two-level atom with ground state $|g\rangle$ and
excited state $|e\rangle$ and  Hamiltonian $H_{A} =
\hbar\omega_{A}|e\rangle\langle e|$. The medium (with fields and bath
included) is described by the damped-polariton model, with Hamiltonian $H_{M}$
given by (\ref{HHB}). The total Hamiltonian is $H = H_0 + V$, with $H_0 =
H_{M} + H_{A}$ and $V=-\bbox{\mu}_{A}\cdot {\bf E}({\bf r}_{A})$, the dipole
interaction between the atom and the medium; $\bbox{\mu}_{A}$ is the atomic
dipole moment operator and ${\bf E}({\bf r}_{A})$ is the electric field
operator at the position ${\bf r}_{A}$ of the atom.

Suppose that the damped-polariton system is prepared at time $0$ in a state
described by a density matrix $\rho_M(0)$. We do not assume that $\rho_M(0)$
commutes with $H_M$, nor that it factorizes into a product of a density
operator for the bath and a density operator for the undamped-polariton system
(as is often assumed for convenience\cite{CohenDG92}). At time $t_{0}>0$ we
bring the guest atom in its excited state, and couple it to the
damped-polariton system. Using perturbation theory, one can calculate
\cite{MandelW95} the time-dependent probability that the guest atom has
emitted a photon at time $t>t_{0}$. We define the derivative of this quantity
as the instantaneous spontaneous emission rate $\Gamma(t)$. It is given as
\begin{equation}\label{Gammat}
\Gamma(t) =
\frac{2}{\hbar^{2}}\mbox{Re}\int_{t_{0}}^{t}\mbox{d}t'\;
e^{i\omega_{A}(t-t')}\mbox{Tr}[\rho_{M}(0) \bbox{\mu}\cdot {\bf E}({\bf r}_A,
t)\bbox{\mu}\cdot {\bf E}({\bf r}_A,t')],
\end{equation}
where $\bbox{\mu}$ is now the dipole transition matrix element of the guest
atom.

If the guest atom is excited a long time after the initial preparation of the
medium, all transient effects in the electric field have damped out. Hence,
the field may be replaced by its long-time limit ${\bf E}_l({\bf r}_A,t)$,
which is given in (\ref{fieldlongtime}). Since ${\bf E}_l$ depends only on the
bath operators at $t=0$, we may write (\ref{Gammat}) in the form:
\begin{equation}\label{Gammat2}
\Gamma(t) =
\frac{2}{\hbar^{2}}\mbox{Re}\int_{t_{0}}^{t}\mbox{d}t'\;
e^{i\omega_{A}(t-t')}\mbox{Tr}_{\rm bath}[\rho_{\rm red}(0) \bbox{\mu}\cdot {\bf E}_l({\bf r}_A,
t)\bbox{\mu}\cdot {\bf E}_l({\bf r}_A,t')].
\end{equation}
Here $\rho_{\rm red}$ is the reduced density matrix obtained by tracing out
the electromagnetic and material degrees of freedom: $\rho_{\rm
red}(0)=\mbox{Tr}_{\rm em,mat} \rho_M(0)$. For the special case that the
initial density matrix $\rho_M(0)$ factorizes, the reduced density matrix is
the bath density matrix $\rho_{\rm bath}(0)$ at $t=0$. In general, the initial
state of the electromagnetic and material degrees of freedom at $t=0$ does not
play a role in the emission rate.

Spontaneous emission in its pure form arises if the reduced density matrix
describes the ground state of the bath. Let us assume this is indeed the case.
Upon inserting (\ref{fieldlongtime}) in (\ref{Gammat2}) we can perform the
$t'$-integral, the integrals over the wavevector and the summations over the
polarization directions. This leads to
\begin{equation}\label{Gammat3}
\Gamma(t)=
\frac{\mu^{2}}{3\pi^{2}\hbar\varepsilon_{0}c^{3}}\mbox{Re}
\int_{0}^{\infty}\mbox{d}\omega
\omega^{3}n(\omega)\frac{\sin[(\omega -
\omega_{A})(t-t_{0})]}{\omega - \omega_{A}},
\end{equation}
with $n(\omega)=\sqrt{\epsilon(\omega)}$ the complex refractive index.

For times $(t-t_{0})$ that are large enough, one may replace $\sin[(\omega -
\omega_{A})(t-t_{0})]/(\omega - \omega_{A})$ by $\pi
\delta(\omega-\omega_{A})$. However, the time scale at which this replacement
is valid, depends on the resonance structure of the refractive index
$n(\omega)$. Since we want to study just this time scale, we will not make the
replacement. To evaluate the integral we multiply the integrand by a
convergence factor $\Omega^{4}/(\Omega^{4} + \omega^{4})$, with
$\Omega\gg\omega_{A}$. The specific choice of the cut-off frequency $\Omega$
will only affect $\Gamma(t)$ at time differences $t-t_{0}$ much smaller than a
single optical cycle. We need to use a high-frequency cut-off at this point,
because the dipole approximation is incorrect for high frequencies.

\begin{figure}
\begin{center}
{\includegraphics[width=85mm, height=60mm]{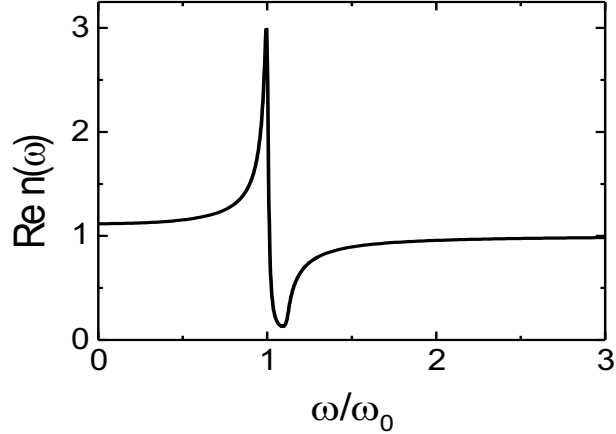}}
\end{center}
\caption{\small Real part of the refractive index in the Lorentz
oscillator model, when $\varepsilon(\omega)$ is given by equation
(\ref{sumepslor}) with parameters $\kappa_{0}=0.01\omega_{0}$,
$\omega_{c}=0.5\omega_{0}$, $\Omega=10\omega_{0}$.}\label{reimn}
\end{figure}

For the dielectric function, we take the Lorentz oscillator form
(\ref{sumepslor}), and we choose the cut-off frequency in that
model to be identical to the one inserted in (\ref{Gammat3}). In
Fig.\ \ref{reimn}, we give the real part of the refractive index
which clearly changes rapidly near $\omega=\omega_{0}$. It is a
familiar figure and it shows that the refractive index does not
change much while increasing the cut-off frequency $\Omega$ from
$10\omega_{0}$ to infinity. The density of radiative modes around
the material resonance is proportional to
$\omega^{2}\mbox{Re}[n(\omega)]$.

With this model for the dielectric function and the parameters as in Fig.\
\ref{reimn}, we calculated $\Gamma(t)$ in the case that the transition
frequency $\omega_{A}$ exactly equals $\omega_{0}$. Since the integrand in
(\ref{Gammat3}) is rapidly fluctuating, it is expedient to use complex
countour deformation to evaluate the integral. We add an infinitesimal
positive imaginary part to the denominator and split the sine into two complex
exponentials. The contour of the integral with
$\exp[i(\omega-\omega_A)(t-t_0)]$ in the integrand is deformed towards the
positive imaginary axis. The contribution from the pole arising from the
convergence factor can be neglected at time scales $t-t_0\gg\omega_{A}^{-1}$.
Likewise, the integration contour of the integral with
$\exp[-i(\omega-\omega_A)(t-t_0)]$ is deformed towards the negative imaginary
axis. Again, the pole contribution from the convergence factor is negligible.
Further contributions, which cannot be neglected, arise from the branch cuts
of $n(\omega)=\sqrt{\varepsilon(\omega)}$ and from the pole at $\omega_{A}$.
The latter contribution is easily evaluated and yields the equilibrium value
$\Gamma(\infty)=\Gamma_0 \; \mbox{Re}\; n(\omega_A)$. In contrast, the branch
cuts yield time-dependent contributions to $\Gamma(t)$. For large $\Omega$
they are situated at $\omega_{1} = -i\kappa_{0} +
\sqrt{\omega_{0}^{2}-\kappa_{0}^{2}}$ and $\omega_{2} = -i\kappa_{0} +
\sqrt{\omega_{0}^{2}+\omega_{c}^{2}-\kappa_{0}^{2}}$. Around $\omega_{1}$ and
$\omega_{2}$, we can approximate the dielectric function by
\begin{eqnarray}
\varepsilon(\omega_{1}-i\delta e^{i\varphi}) & \simeq &
-i\omega_{c}^{2}
e^{-i\varphi}/(2\delta\sqrt{\omega_{0}^{2}-\kappa_{0}^{2}}),
\\
\varepsilon(\omega_{2}-i\delta e^{i\varphi}) &
\simeq & -2 i\delta e^{i\varphi}
\sqrt{\omega_{0}^{2}+\omega_{c}^{2}-\kappa_{0}^{2}}
/\omega_{c}^{2}.
\end{eqnarray}
The branch cut at $\omega_{1}$ gives the following contribution to the
spontaneous emission rate:
\begin{equation}\label{analytical1}
-\frac{\Gamma_{0}}{\pi\omega_A^3}\mbox{Re}\left[
e^{-i\pi/4-\kappa_{0}(t-t_{0})
-i(\sqrt{\omega_{0}^{2}-\kappa_{0}^{2}}-\omega_{A})(t-t_{0})}J(t)\right],
\end{equation}
where $J(t)$ is defined as:
\begin{equation}\label{jt}
J(t)=\int_{0}^{\infty}\mbox{d}\lambda\;\frac{e^{-\lambda
(t-t_{0})}[\sqrt{\omega_{0}^{2}-\kappa_{0}^{2}}-i(\lambda+\kappa_0)]^3}
{\sqrt{\lambda}[\sqrt{\omega_{0}^{2}-\kappa_{0}^{2}}-\omega_{A}-i(\lambda
+\kappa_{0})]}\;
\left[\frac{2\omega_c^2\sqrt{\omega_{0}^{2}-\kappa_{0}^{2}}
+i\lambda(\lambda^2+4\omega_0^2+\omega_c^2-4\kappa_0^2)}
{4(\omega_0^2-\kappa_0^2)+\lambda^2}\right]^{1/2}.
\end{equation}
The branch cut around $\omega_2$ gives a similar contribution.

\begin{figure}
\begin{center}
{\includegraphics[width=85mm, height=60mm]{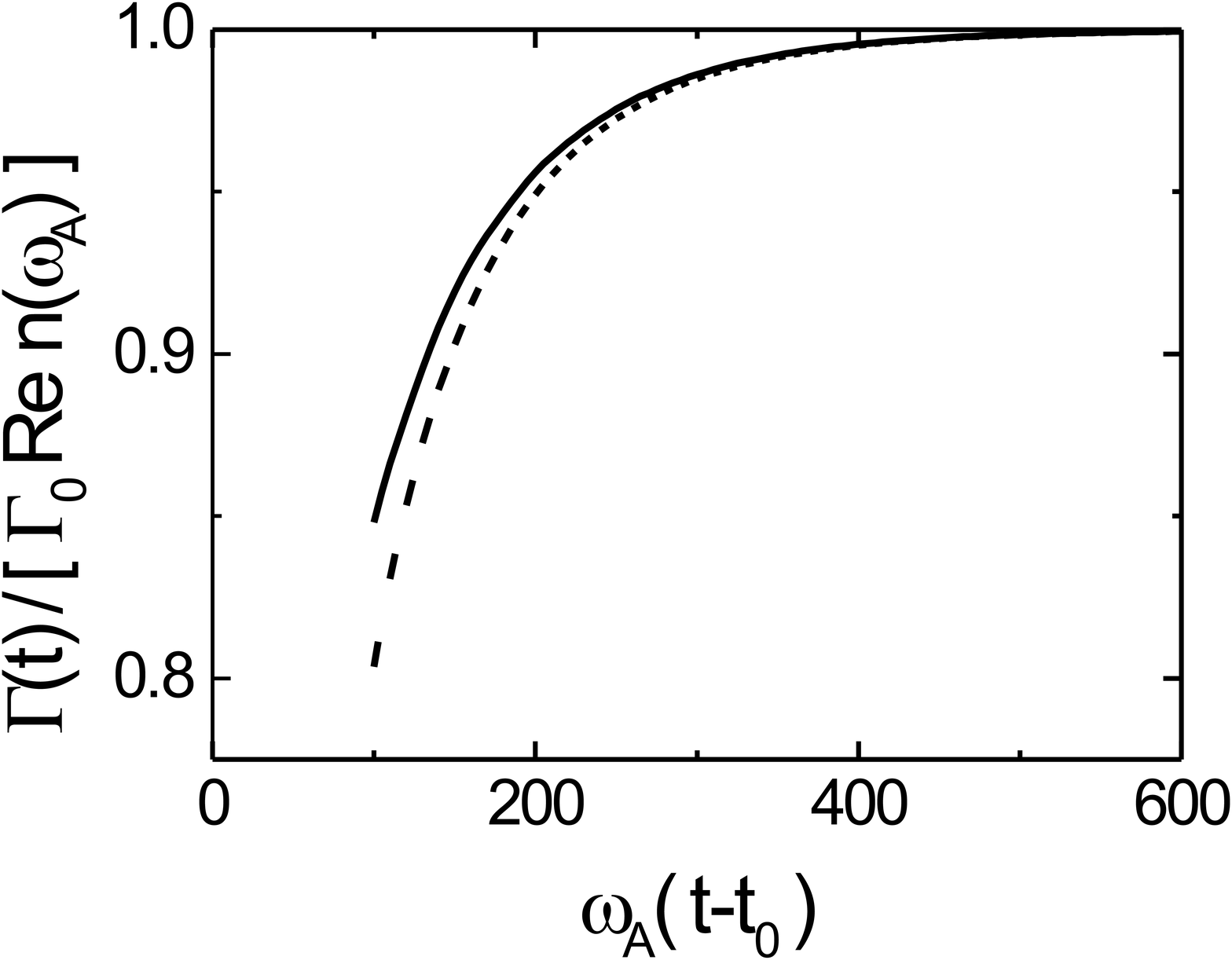}}
\end{center}
\caption{\small Normalized spontaneous emission rate
$\Gamma(t)/[\Gamma_{0}\, \mbox{Re}\,n(\omega_{A})]$  in an
absorbing dielectric as a function of time, when the transition
frequency $\omega_{A}$ is equal to $\omega_{0}$. Choice of
parameters in the Lorentz oscillator model:
$\kappa_{0}=0.01\omega_{0}$, $\omega_{c}=0.5\omega_{0}$, $\Omega=
\infty$.  The solid line is the exact result for (\ref{Gammat3})
and the dashed line is the approximate expression
(\ref{analyticalapp}).}\label{art1fig4}
\end{figure}

The integrals arising from the branch cuts and from the imaginary axis can
easily be evaluated numerically, since their integrands are no longer rapidly
fluctuating. The result is the solid line in Fig.\ \ref{art1fig4}. We see that
the spontaneous emission rate builds up until it finally reaches the
time-independent equilibrium value $\Gamma_{0}\,\mbox{Re}\,n(\omega_{A})$.

The dashed line in Fig.\ \ref{art1fig4} is an analytical approximation for
$\Gamma(t)$, which captures the main features of the time dependence, at least
qualitatively. It is derived by retaining only the contribution
(\ref{analytical1}) in the time dependent part of $\Gamma(t)$, as this is
dominant for large $t$. Moreover, we approximate $J(t)$ by the first term in
its asymptotic expansion for large $t-t_0$. In this way we arrive at the
following approximate expression for $\Gamma(t)$:
\begin{equation}\label{analyticalapp}
\Gamma(t) \simeq \Gamma_{0}\mbox{Re}\left[n(\omega_{A}) -
\frac{\omega_{c}(\omega_{0}^{2}-\kappa_{0}^{2})^{5/4}}
{\sqrt{2\pi}\omega_{A}^{3}
(\sqrt{\omega_{0}^{2}-\kappa_{0}^{2}}-\omega_{A}-i\kappa_{0})}
\frac{e^{-i\pi/4-\kappa_{0}(t-t_{0})
-i[\sqrt{\omega_{0}^{2}-\kappa_{0}^{2}}-\omega_{A}](t-t_{0})}}
{(t-t_{0})^{1/2}}\right].
\end{equation}
As explained, this approximation contains only the contribution from the
branch cut at $\omega_{1}$; the branch cut at $\omega_{2}$ gives a faster
decaying term, which goes like $e^{-\kappa_{0}(t-t_{0})}/(t-t_{0})^{3/2}$. The
contributions from the integrals along the imaginary axis decay even faster.

It can be seen from equation (\ref{analyticalapp}) that the amplitude of the
time-dependent part of $\Gamma(t)$ falls off as
$e^{-\kappa_{0}(t-t_{0})}/(t-t_{0})^{1/2}$ and also, that the amplitude of the
extra term is largest around resonance, when
$\omega_{A}\simeq\sqrt{\omega_{0}^{2}-\kappa_{0}^{2}}$. Away from  resonance
oscillations with frequency $\sqrt{\omega_{0}^{2}-\kappa_{0}^{2}}-\omega_{A}$
are present. Fig.\ \ref{art1fig4} shows the on-resonance case when the
time-dependent term shows no oscillations, but has relatively large amplitude.

The main result of the present discussion is the time-dependence of the
spontaneous emission rate. The time-independent value is not reached
instantaneously, but at a time scale that is governed by the resonance
characteristics of the medium. In fact, the smaller the resonance width
$\kappa_{0}$, the longer it takes to reach the time-independent value.
Typically, it takes $\omega_{0}/\kappa_{0}$ optical cycles, as follows
from the exponential $e^{-\kappa_{0}(t-t_0)}$ in the approximate expression
(\ref{analyticalapp}). For narrow resonances with $\omega_{0}/\kappa_{0}$
large, the transient dynamics may take a substantial amount of time.

\section{Discussion and conclusions}\label{conc}

We have solved the equations of motion for the field operators in the damped
polariton model using Laplace transformations. The solutions of the field and
medium operators are the sum of a transient and a permanent part. The latter
are expressed solely in terms of the initial bath operators. Long after the
initial time all field and medium operators are functions of the bath
operators alone, provided the coupling to the bath is nonzero for all
frequencies. The long-time solutions satisfy quantum Langevin equations in
which the initial bath operators figure as the quantum noise source. The same
continuum that produces the absorption also forms the noise source that keeps
the commutation relations in order. This is conceptually simpler than
expressing the quantum Langevin noise in terms of the creation and
annihilation operators that diagonalize the total Hamiltonian of the
damped-polariton model\cite{HuttnerB92b}.

The effects of the initial state of the field and medium variables on the
expectation values at a later time are noticeable only during a short period
that is determined by the characteristic relaxation times of the damped
polariton modes. Once these transient effects have died out the expectation
values are determined by the reduced density matrix which follows from the
full density matrix at the initial time by taking the trace over the degrees
of freedom of field and matter (without bath). If the full density
matrix at the initial time factorizes, the reduced density matrix equals
the initial bath density matrix.

The method of long-time solutions can be used for other dissipative quantum
systems as well. For models in which the Hamiltonian can be diagonalized
completely, it is an alternative to the Fano diagonalization technique. The
latter can be quite complicated\cite{HuttnerB92b,RosenauCDW00}, whereas our
long-time solutions are found after the simple inversion of a $4\times 4$
matrix, as one sees from section \ref{model} and \ref{longtime}. More
generally, the long-time method may be useful for dissipative systems with a
bilinear coupling to a harmonic oscillator bath whose dynamics can be
integrated out.

We employed the method of long-time solutions to study transient
effects in a medium described by a Lorentz oscillator dielectric
function. This dielectric function (and that of the
point-scattering model as well) can be derived from the
damped-polariton model by taking a suitable bath coupling.
Although a cut-off procedure turns out to be indispensable, the
essential physics in the optical regime can be represented
adequately in this way. Once the connection with the
damped-polariton model has been established, spontaneous emission
processes by a guest atom in a Lorentz oscillator dielectric can
be investigated by means of the long-time method. Although
transient effects due to the initial preparation of the dielectric
have damped out after a few medium relaxation periods, transient
behaviour of a different type shows up in the initial stages of
the decay process. This transient behaviour, which is related to
the preparation of the guest atom in its excited state, leads to a
non-exponential decay - or in other words to a time-dependent
spontaneous emission rate - if the atomic transition frequency is
near a resonance of the dielectric. The non-exponential dynamics
takes place at time scales that are inversely proportional to the
width of the resonance. As we have shown, the characteristics of
the time-dependent decay rate can be captured in an analytic
asymptotic expression of which the qualitative features are
corroborated by numerical methods.

\section*{Acknowledgements}
We would like to thank Ad Lagendijk, Rudolf Sprik and Willem Vos for 
stimulating discussions. This work is part of the research program of the 
Stichting 
voor Fundamenteel Onderzoek der Materie, which is financially supported
by the Nederlandse Organisatie voor Wetenschappelijk Onderzoek.

\appendix

\section*{Laplace operators and time-dependent coefficients}\label{appA}

In section \ref{model} the electric field operator $\bar E(p)$ was given in
terms of the operators at $t=0$. Here we give the analogous expressions for
the other Laplace operators. Furthermore, we show how to evaluate the
time-dependent coefficients $M_{mn}(t)$ like in (\ref{timeco2}) for the
electric field operator. Finally, we list the expressions for the coefficients
of the other operators.

The expression for $\bar E(p)$ in (\ref{Einfields0}) has the following
analogous expressions for the other Laplace operators:
\begin{eqnarray}\label{allinfields0}
 \bar A(p)& = & \frac{1}{\bar D(p)}
\left\{\rule{0mm}{5mm} -E(0) +p A(0)\right. \nonumber \\
& &\left.-\frac{\alpha}{\varepsilon_{0}}
\left[\frac{p^{2}}{\omega_{c}^{2}}\bar\chi(p)
-1\right]X(0) -\frac{1}{\alpha}p\bar\chi(p)[P(0)-\bar B(p)] \right\}, \\
\bar X(p)& = & \frac{1}{\bar D(p)} \left\{
-\frac{\varepsilon_{0}}{\alpha} p \bar\chi(p) E(0)
+\frac{\varepsilon_{0}}{\alpha}p^{2}\bar\chi(p)A(0) \right.\nonumber\\
 & & \left. +
\left(\frac{k^{2}c^{2}}{\omega_{c}^{2}} +\frac{p^{2}}{\omega_{c}^{2}}
+1\right)p\bar\chi(p)X(0)  + \frac{\varepsilon_{0}}{\alpha^{2}}(p^{2}
+k^{2}c^{2})\bar\chi(p)[P(0) - \bar B(p)] \right\}, \\
\bar P(p)&=&\frac{1}{\bar D(p)} \left\{
-\alpha\left[\frac{p^{2}}{\omega_{c}^{2}}\bar\chi(p)-1\right]E(0) +\alpha p
\left[\frac{p^{2}}{\omega_{c}^{2}}\bar\chi(p)-1\right]A(0)\right. \nonumber \\
& &\left. \quad +
\frac{\alpha^{2}}{\varepsilon_{0}}\left[\frac{p^{2}}{\omega_{c}^{2}}
\bar\chi(p)-1\right]
\left(\frac{k^{2}c^{2}}{\omega_{c}^{2}}+\frac{p^{2}}{\omega_{c}^{2}} +1\right)
X(0)\right. \nonumber \\
& &\left.+p\left(\frac{k^{2}c^{2}}{\omega_{c}^{2}}
+\frac{p^{2}}{\omega_{c}^{2}} +1\right)\bar\chi(p)[P(0)-\bar B(p)] \right\}.
\end{eqnarray}
If we now apply the inverse Laplace transformation to these expressions, we
find the full time dependence of the operators $A$, $X$ and $P$. The inverse
Laplace transformation is a contour integration over the Bromwich contour that
includes the whole imaginary $p$-axis. After transforming to frequency
variables, the contour includes poles from $\bar D^{-1}(p)$, which are in the
lower halfplane, and moreover poles on the real frequency axis arising from
$\bar B(p)$. The latter are important in the calculation of the long-time
solutions of the operators in section \ref{longtime}. However, in the
calculation of the coefficients $M_{mn}(t)$, which we will discuss here, they
play no role. Let us consider as an example the coefficient $M_{AE}(t)$:
\begin{eqnarray}\label{meaexample}
M_{AE}(t) &  = & - \frac{1}{2\pi i}
\int_{-i\infty+\eta}^{i\infty+\eta}\mbox{d}p\;e^{pt}\bar D^{-1}(p)=
\frac{1}{2\pi}\int_{-\infty}^{\infty}\mbox{d}\omega\frac{e^{-i\omega
t}}{\varepsilon(\omega)\omega^{2}-k^{2}c^{2}} \nonumber \\
& = & \frac{1}{4\pi k c}\int_{-\infty}^{\infty}
\mbox{d}\omega\left( \frac{e^{-i\omega
t}}{n(\omega)\omega -k c} - \frac{e^{-i\omega t}}{n(\omega)\omega
+k c}\right) = \frac{1}{k
c}\sum_{j}\mbox{Im}(e^{-i\Omega_{j}t}\frac{v_{g,j}}{c}).
\end{eqnarray}
Note that $M_{AE}(t)$ is exponentially damped because all
$\Omega_{j}$ in the exponentials have negative imaginary parts.
The other coefficients can be calculated in a similar way. The
results are:
\begin{eqnarray}\label{tcoeffall}
M_{AA}(t) & = & M_{EE}(t), \\
M_{AX}(t) & = & -\frac{\alpha k}{\omega_{c}^{2}\varepsilon_{0}c^{4}}
\sum_{j}\mbox{Im}\;\left[e^{-i\Omega_{j}t}v_{g,j}(1-\frac{v_{p,j}^{2}}{c^{2}}
+\frac{\omega_{c}^{2}}{k^{2}c^{2}})\right], \\
M_{AP}(t) & = &\frac{1}{\alpha} \sum_{j}\mbox{Re}\;
\left[e^{-i\Omega_{j}t}\frac{v_{g,j}}{c}\left( \frac{v_{p,j}}{c} -
\frac{c}{v_{p,j}}\right)\right], \\
M_{XE}(t) & = & \varepsilon_{0}M_{AP}(t), \\
M_{XA}(t) & = & \varepsilon_{0}M_{EP}(t),\\
M_{XX}(t) & = & -\frac{k^{2}c^{2}}{\omega_{c}^{2}}\sum_{j}\mbox{Re}\;
\left[e^{-i\Omega_{j}t}\frac{v_{g,j}}{c}\left(\frac{v_{p,j}}{c} -
\frac{c}{v_{p,j}}\right)\left(1-\frac{v_{p,j}^{2}}{c^{2}} +
\frac{\omega_{c}^{2}}{k^{2}c^{2}}\right)\right], \\
M_{XP}(t) & = & -\frac{\varepsilon_{0}k c}{\alpha^{2}}\sum_{j}\mbox{Im}\;
\left[e^{-i\Omega_{j}t}\frac{v_{g,j}}{c}\left(\frac{v_{p,j}}{c}
- \frac{c}{v_{p,j}} \right)^2\right], \\
M_{PE}(t) & = & \varepsilon_{0}M_{AX}(t),\\
M_{PA}(t) & = & \varepsilon_{0} M_{EX}(t), \\
M_{PX}(t) & = &
\frac{\alpha^2 k^{3}c^{3}}{\omega_{c}^{4}\varepsilon_{0}}\sum_{j}\mbox{Im}\;
\left[e^{-i\Omega_{j}t}\frac{v_{g,j}}{c}\left(1-\frac{v_{p,j}^{2}}{c^{2}}+
\frac{\omega_{c}^{2}}{k^{2}c^{2}} \right)^{2}\right], \\
M_{PP}(t)& = & M_{XX}(t).
\end{eqnarray}
With the sum rules discussed in section \ref{sumrules}, one can
see that the ``diagonal''  coefficients in this list equal $1$ at
time $t=0$, whereas the other coefficients have the initial value
$0$.

\end{document}